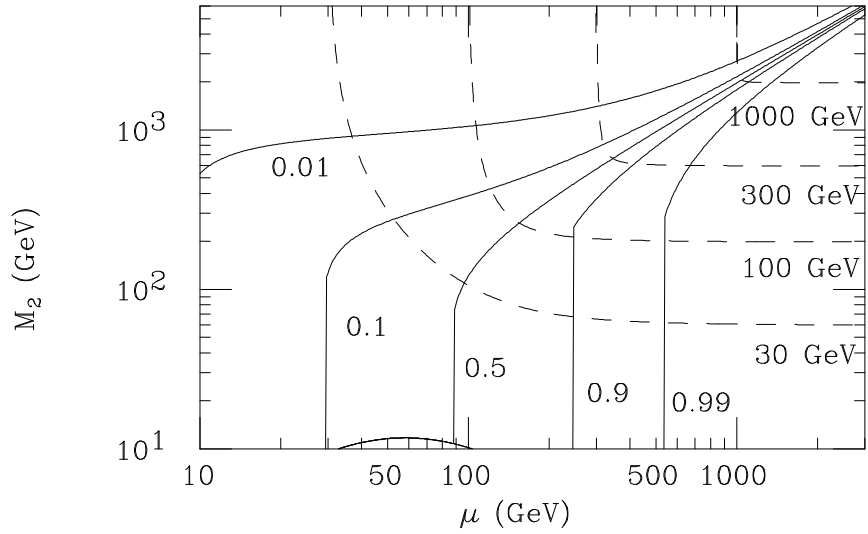

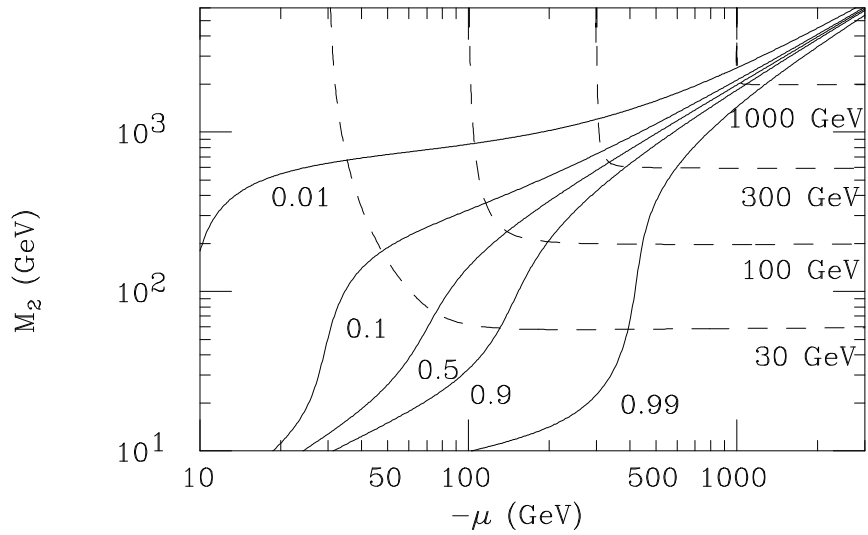

**Figure 1.** Iso–mass and iso–composition lines for neutralino in the $M_2$–$\mu$ plane for $\tan\beta = 8$. Dashed lines are lines of constant $m_\chi$; solid lines are lines of constant gaugino fractional weight $P$. The upper figure refers to the case $\mu > 0$, the lower to $\mu < 0$.

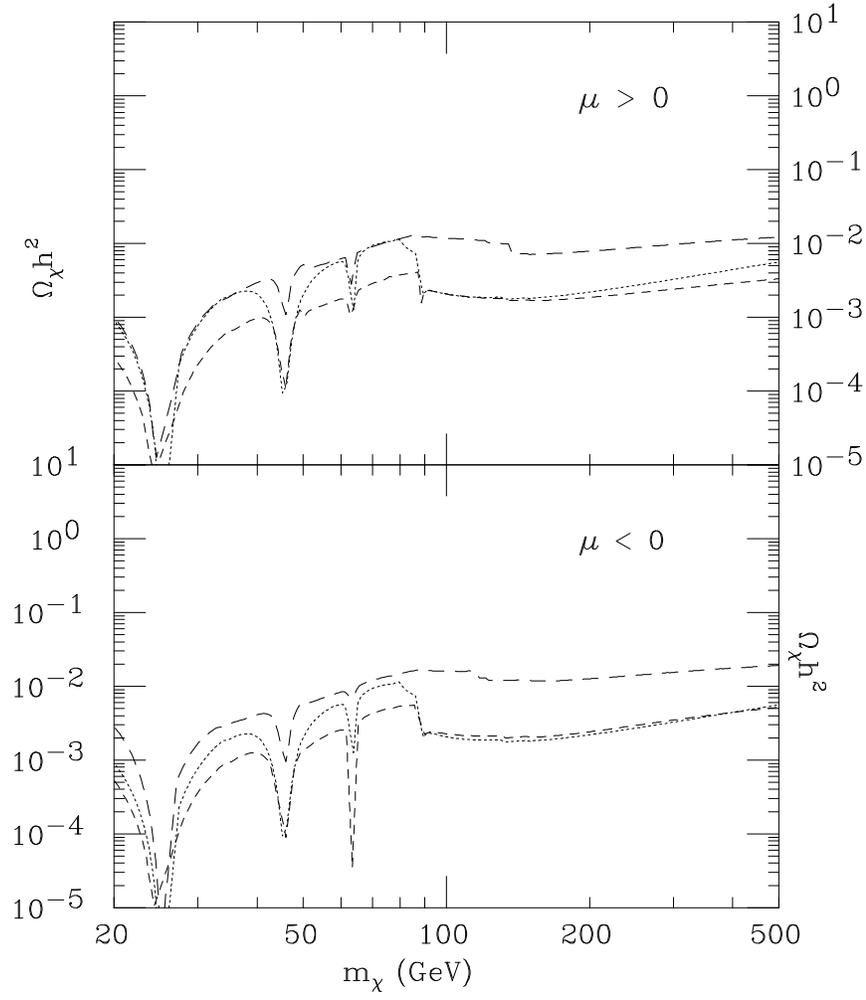

**Figure 2.** Neutralino relic abundance $\Omega_\chi h^2$ as a function of $m_\chi$ for the three representative neutralino compositions $P = 0.1$ (dotted line), 0.5 (short–dashed line), 0.9 (long–dashed line).

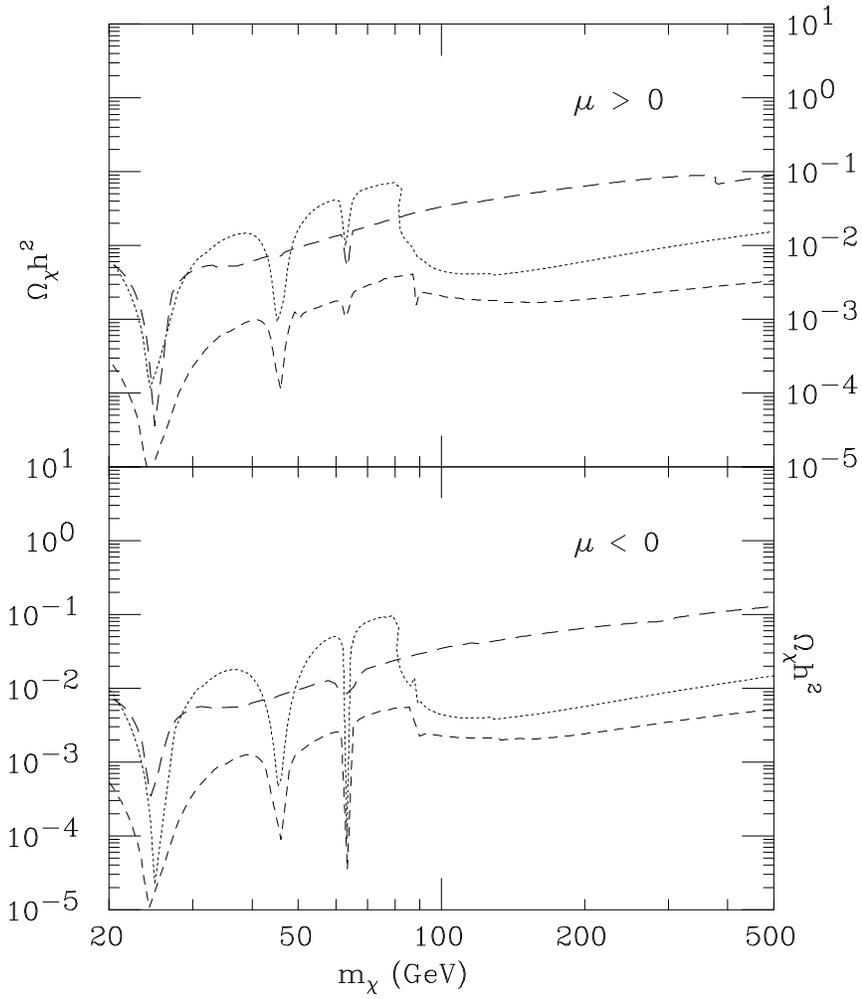

**Figure 3.** Same as in Fig.2, except that now composition are: $P = 0.01$ (dotted line), 0.5 (short–dashed line), 0.99 (long–dashed line).

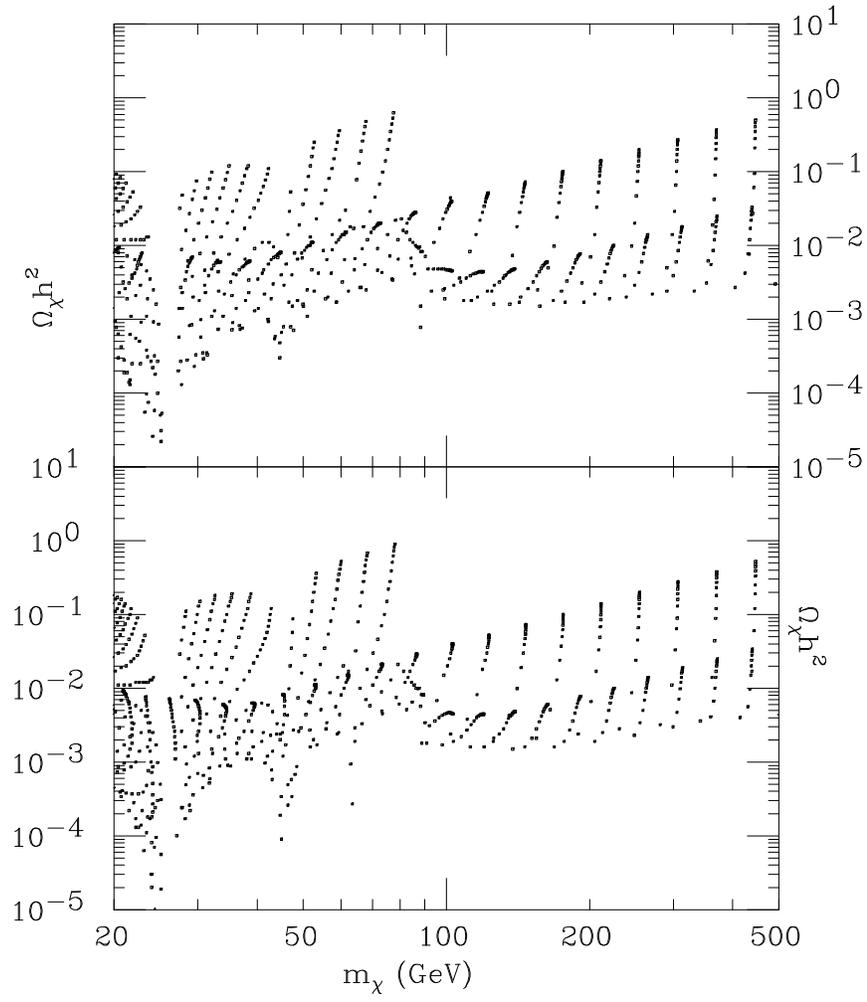

**Figure 4.** Scatter plot for the neutralino relic abundance $\Omega_\chi h^2$ as a function of $m_\chi$.

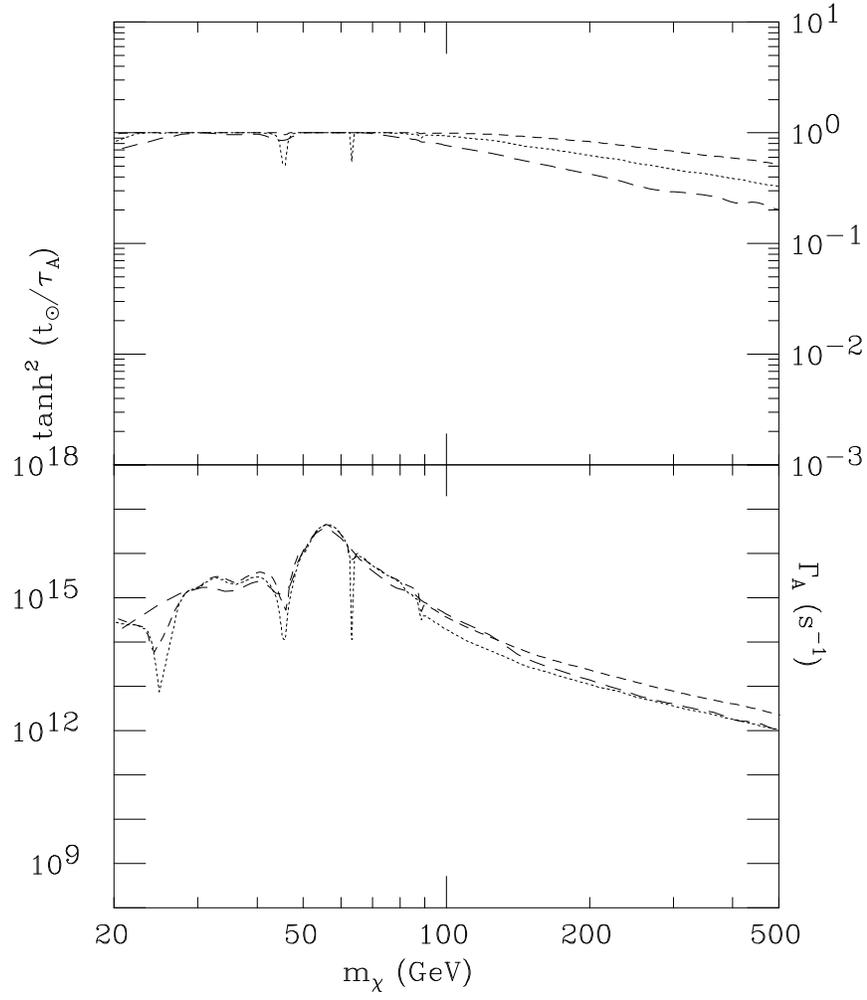

**Figure 5.** $\tanh^2(t_\odot/\tau_A)$ and $\Gamma_A$ are given as functions of $m_\chi$ for positive values of the parameter $\mu$, for the three representative neutralino compositions $P = 0.1$ (dotted line), 0.5 (short–dashed line), 0.9 (long–dashed line). This figure refers to the Earth.

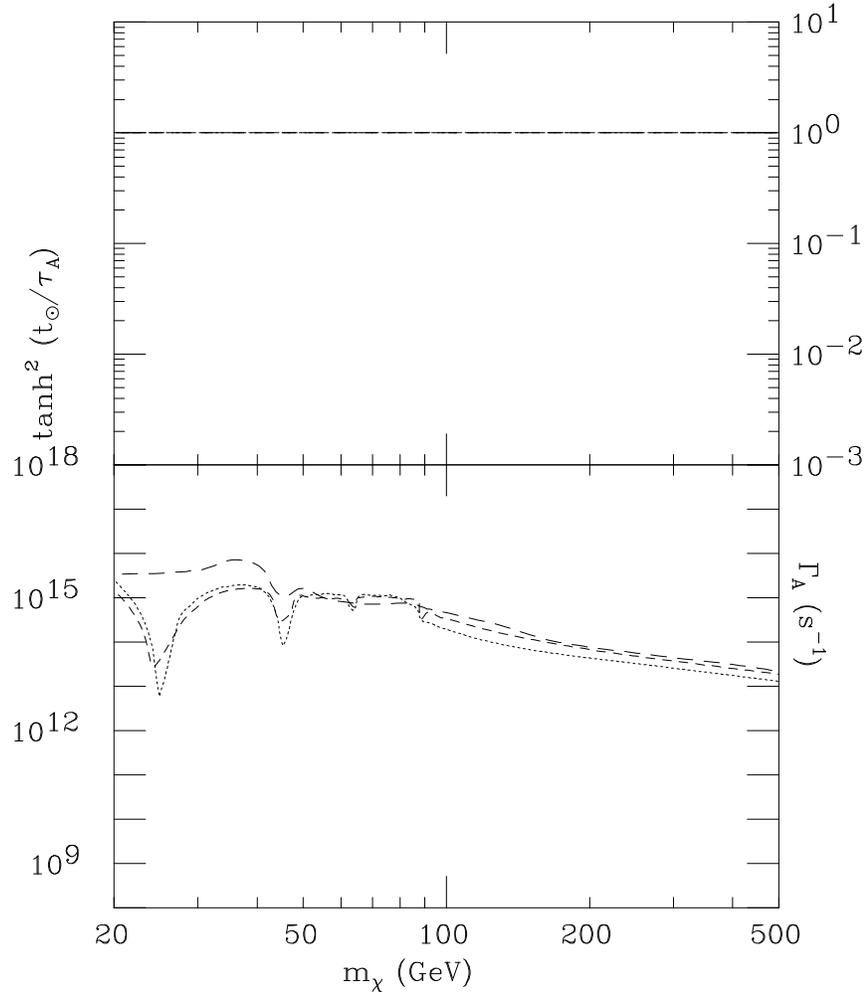

**Figure 6.** Same as in Fig.5 for the case of the Sun. To make the comparison of the annihilation rates for the Earth and for the Sun easier, the plot shown here for the Sun refers to the effective annihilation rate, defined as $\Gamma_A$ times the ratio $(d_1/d_2)^2 = 1.8 \times 10^{-9}$, where $d_1$ is the radius of the Earth and $d_2$ is the Sun-Earth distance.

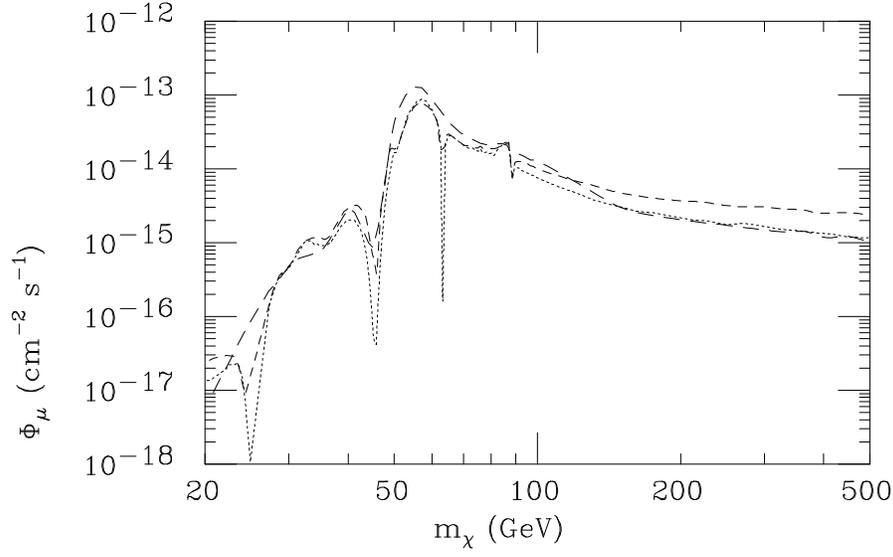

**Figure 7.** Fluxes of the up–going muons as functions of $m_\chi$ for $\chi$–$\chi$ annihilation in the Earth, for the three representative neutralino compositions $P = 0.1$ (dotted line), 0.5 (short–dashed line), 0.9 (long–dashed line). The threshold for the muon energy is $E_\mu^{\text{th}} = 2$ GeV.

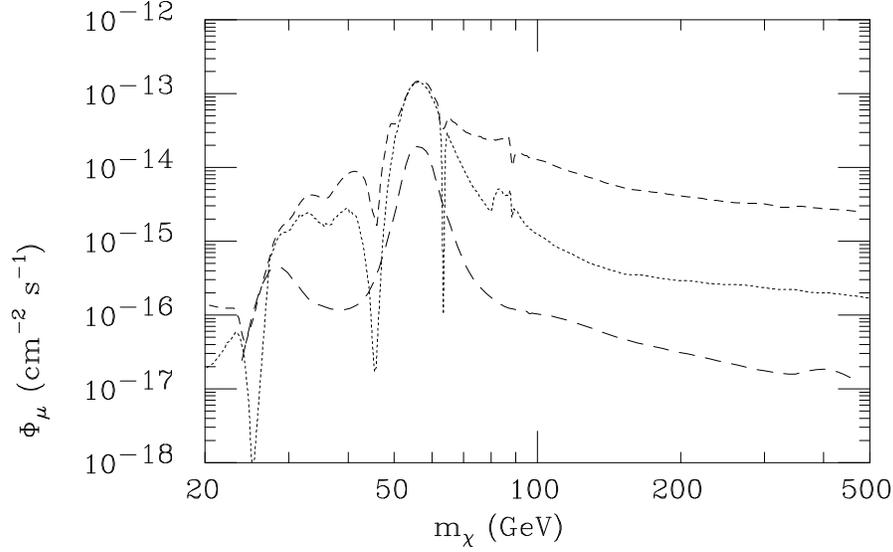

**Figure 8.** Same as in Fig.7 except now the compositions are: $P = 0.01$ (dotted line), 0.5 (short–dashed line), 0.99 (long–dashed line).

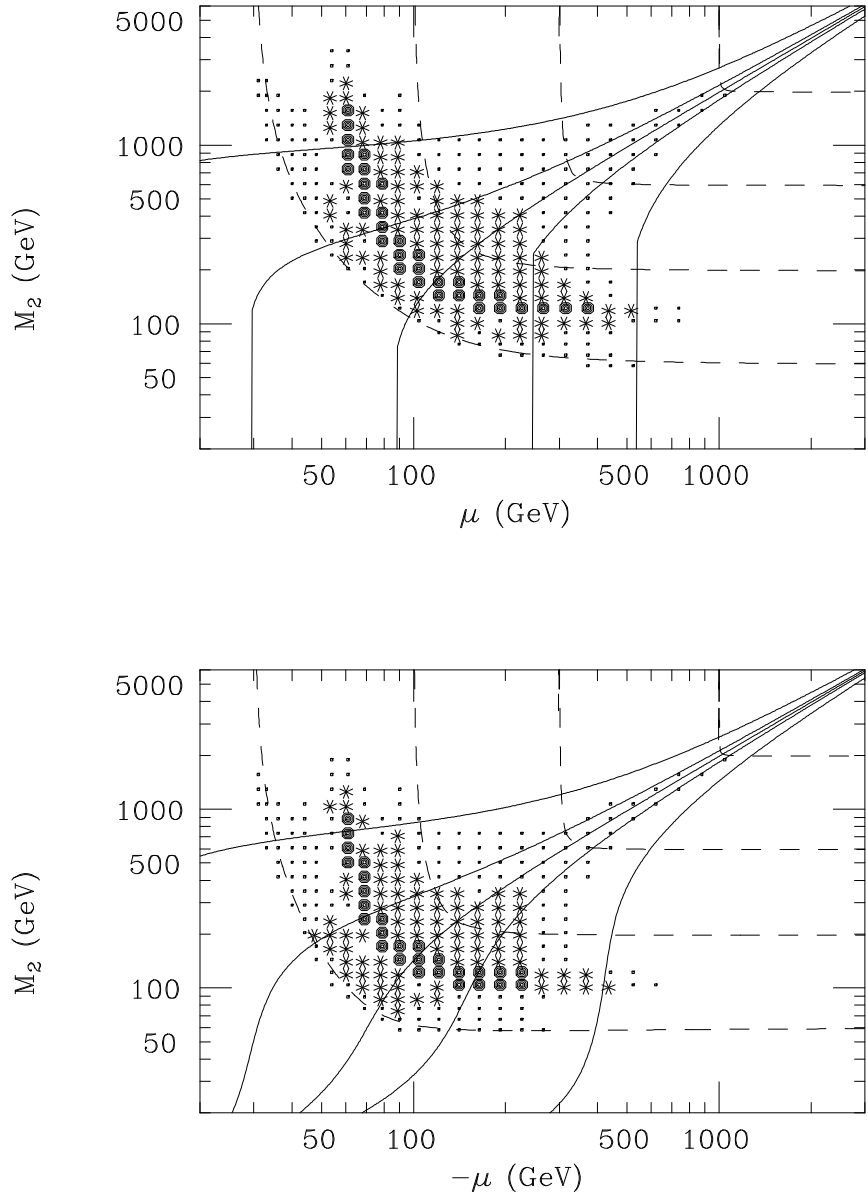

**Figure 9.** In the $M_2$–$\mu$ plane are shown the regions explored by Kamiokande (full circles) and the domains explorable by an improvement factor of 10 (asterisks) and of 100 (dots) in sensitivity. Dashed lines denote iso–mass curves with $m_\chi = 30, 100, 300, 1000$ GeV. Solid lines denote curves of iso–composition (here $P = 0.01, 0.1, 0.5, 0.9, 0.99$ are shown).

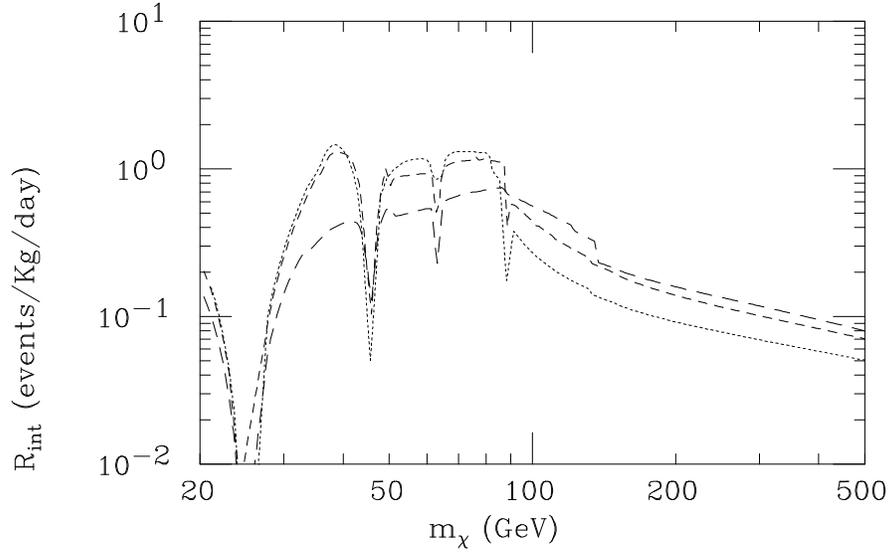

**Figure 11.** Event rates for detection of neutralino dark matter for a (natural composition) Germanium detector for the three representative neutralino compositions $P = 0.1$ (dotted line), 0.5 (short–dashed line), 0.9 (long–dashed line). The rate refers to the events with an electron–equivalent energy in the range (2–4) KeV.

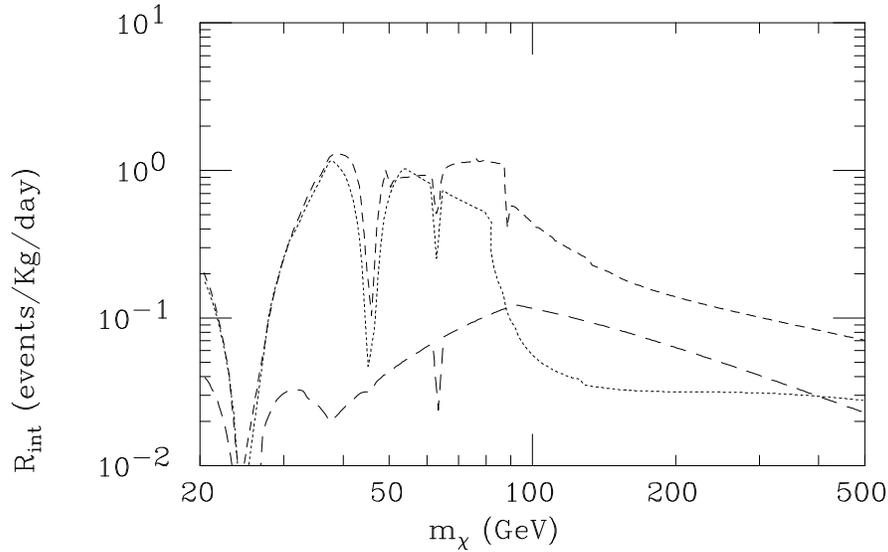

**Figure 12.** Same as in Fig.11 except that now the three representative neutralino compositions are $P = 0.01$ (dotted line), 0.5 (short–dashed line), 0.99 (long–dashed line).

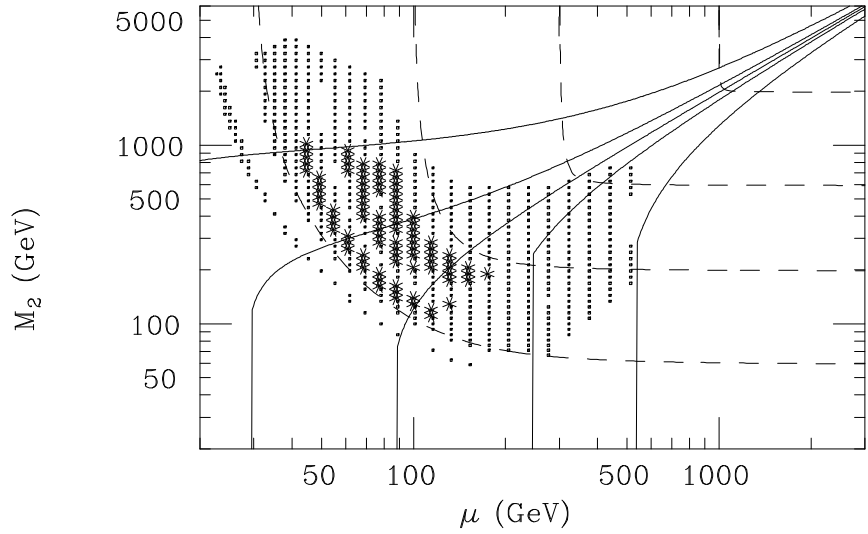

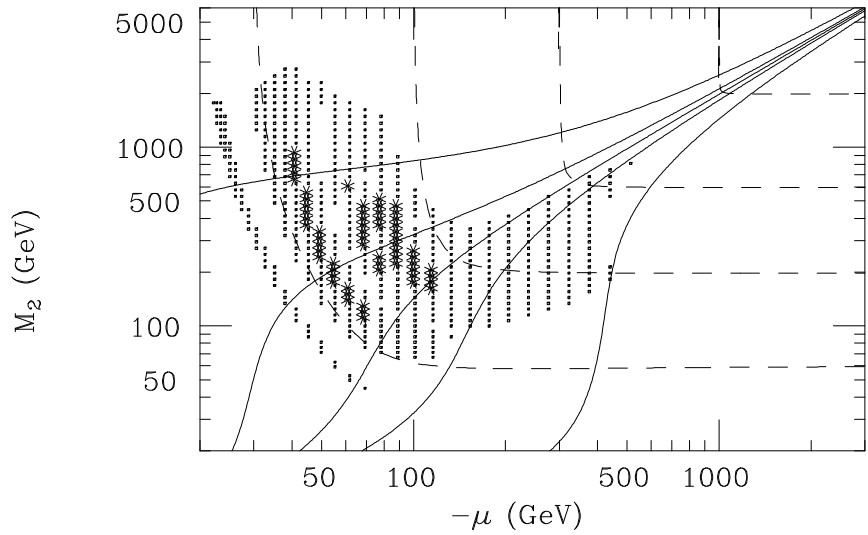

$\tan\beta = 8$

**Figure 13.** Explorable regions in the $M_2$–$\mu$ plane for neutralino–Ge interaction. Asterisks denote regions where the integrated rate $R_{\rm int}$ is in the range $0.1 R_{\rm exp} < R_{\rm int} \leq R_{\rm exp}$, $R_{\rm exp}$ being the present experimental bound. Dots denote regions where the integrated rate $R_{\rm int}$ is in the range $0.01 R_{\rm exp} < R_{\rm int} \leq 0.1 R_{\rm exp}$,

# NEUTRALINOS AS DARK MATTER CANDIDATES

A. BOTTINO, N. FORNENGO, G. MIGNOLA

*Dipartimento di Fisica Teorica, Università di Torino and*
*INFN, Sezione di Torino, via P.Giuria 1, 10125 Torino, Italy*

and

S. SCOPEL

*Dipartimento di Fisica, Università di Genova and*
*INFN, Sezione di Genova, via Dodecaneso 33, 16146 Genova, Italy*

ABSTRACT

We review some properties of the neutralino as a candidate for dark matter in the Universe. After presentation of evaluations for the neutralino relic abundance, possibilities for its direct and indirect detections are discussed, with emphasis for measurements at neutrino telescopes.

## 1  The cold side of the dark: neutralinos

As is widely discussed in the literature, neutralino may be considered as a very natural candidate for Cold Dark Matter (CDM). This property rests on two assumptions, i.e. that: i) R-parity is conserved, ii) neutralino is the lightest supersymmetric particle (LSP). Very convenient theoretical frameworks where dark matter neutralino phenomenology may be easily studied are provided by the Minimal Supersymmetric Standard Model (MSSM) and by its implementation in a Supergravity theory (SUGRA) [1-2].

The neutralino ($\chi$) is defined as the lowest–mass linear combination of photino, zino and higgsinos

$$\chi = a_1 \tilde{\gamma} + a_2 \tilde{Z} + a_3 \tilde{H}_1^0 + a_4 \tilde{H}_2^0 \ . \tag{1}$$

---

Invited talk presented by A.Bottino at the 6[th] International Workshop on: "Neutrino Telescopes", Venice 1994

Here $\tilde{\gamma}$ and $\tilde{Z}$ are linear combinations of the U(1) and SU(2) neutral gauginos, $\tilde{B}$ and $\tilde{W}_3$,

$$\begin{aligned}\tilde{\gamma} &= \cos\theta_W \tilde{B} + \sin\theta_W \tilde{W}_3, \\ \tilde{Z} &= -\sin\theta_W \tilde{B} + \cos\theta_W \tilde{W}_3,\end{aligned} \quad (2)$$

and $\theta_W$ is the Weinberg angle.

The neutralino mass $m_\chi$ and the coefficients $a_i$ depend on the parameters: $\mu$ (Higgs mixing parameter), $M_1$, $M_2$ (masses of $\tilde{B}$ and of $\tilde{W}_3$, respectively) and $\tan\beta = v_u/v_d$ ($v_u$ and $v_d$ are the v.e.v.'s which give masses to up–type and down–type quarks). It is customary to employ the standard GUT relationship between $M_1$ and $M_2$: $M_1 = (5/3)\tan^2\theta_W M_2 \simeq 0.5\ M_2$. We use this assumption here.

In the following for the parameters $M_2$ and $\mu$ we will consider the ranges: 20 GeV $\leq M_2 \leq$ 6 TeV, 20 GeV $\leq |\mu| \leq$ 3 TeV. tan$\beta$ will be taken at the representative value tan$\beta = 8$.

For the evaluation of the neutralino relic abundance and of the event rates for direct and indirect neutralino detections one also has to assign values to the masses of a large number of particles, namely to the Higgs bosons and to the Susy scalar partners of leptons and quarks: sleptons ($\tilde{l}$) and squarks ($\tilde{q}$). In the MSSM scheme these values are assigned arbitrarily: a standard procedure consists in assuming mass degeneracy both for sleptons and for squarks. Only for stop particles of different helicities non-degeneracy is sometimes introduced, since, under certain circumstances, stop-mixing may generate sizeable effects [3]. As for the neutral Higgs bosons we recall that in the MSSM there are three neutral Higgs particles: two CP–even bosons $h$ and $H$ (of masses $m_h$, $m_H$ with $m_H > m_h$) and a CP–odd one $A$ (of mass $m_A$). Once a value for one of these masses (say, $m_h$) is assigned, the other two masses ($m_A$, $m_H$) are derived through mass relationships depending on radiative effects.

Implementation of MSSM with supergravity sets a much more constrained phenomenological framework, since SUGRA establishes strict relations between all the masses in play and the few fundamental theoretical parameters: $A$ and $B$ (appearing in the soft symmetry-breaking interaction terms), $m_0$ (common scalar mass at the GUT scale), $m_{1/2}$ (common gaugino mass at the GUT scale) and $\mu$. Furthermore, other specific theoretical requirements (features of the symmetry breaking, condition that the neutralino be the LSP, ...) strongly restrict the whole parameter space. This has important consequences; for instance, it constrains the neutralino to compositions with dominance of the gaugino components.

In the present note we wish to discuss the event rates for neutralino searches in a framework which is not too much constrained by theoretical requirements; we then simply adopt here a MSSM, with choices for the free parameters that are only restricted by experimental bounds. Our main concern is to discuss the minimal sensitivity required in experimental devices in order to undertake a significant investigation of neutralino dark matter. For this reason we present here evaluations where the unknown masses are assigned the smallest values compatible with experimental lower bounds; this usually provides maximal values for the signals. To be definite, in the following we will set the sfermion masses at the value $m_{\tilde{f}} = 1.2\ m_\chi$, when $m_\chi > 45$ GeV, $m_{\tilde{f}} = 45$ GeV otherwise. Only the mass of the stop quark is assigned a larger value of 3 TeV. Higgs mass $m_h$ is set at the value of 50 GeV. The top mass has been fixed at $m_t = 150$ GeV.

We show in the $M_2 - \mu$ plots of Fig.1 the iso–mass lines and the iso–composition lines for the neutralino. Along an iso–composition line the composition parameter $P$, defined as the gaugino fractional weight, i.e. $P = a_1{}^2 + a_2{}^2$, is kept fixed. Shown in Fig.1 are the iso–composition lines referring to the values $P = 0.01, 0.1, 0.5, 0.9, 0.99$ (i.e., from a very pure higgsino composition to a very pure gaugino composition).

In the following we will characterize a neutralino state by the values of $m_\chi$ and $P$; we will accordingly present our results for the relic abundance and the event rates in terms of these two parameters.

## 2   How many neutralinos around us?

For the computation of the direct and indirect event rates for neutralino one has to use a specific value for the neutralino density $\rho_\chi$. Obviously, it would be inappropriate to assign to the neutralino local density $\rho_\chi$ the standard value for the total dark matter density $\rho_l = 0.3$ GeV cm$^{-3}$, unless one specifically verifies that the neutralino relic abundance $\Omega_\chi h^2$ turns out to be at the level of an $(\Omega h^2)_{\text{min}}$ consistent with $\rho_l$. This is why a correct evaluation of the event rates for $\chi$ detection also requires a calculation of its relic abundance.

Thus we evaluate $\Omega_\chi h^2$ and we determine $\rho_\chi$ by adopting a standard procedure [4]: when $\Omega_\chi h^2 \geq (\Omega h^2)_{\text{min}}$, we put $\rho_\chi = \rho_l$; when $\Omega_\chi h^2$ turns out to be less than $(\Omega h^2)_{\text{min}}$, we take

$$\rho_\chi = \rho_l \frac{\Omega_\chi h^2}{(\Omega h^2)_{\text{min}}} \ . \tag{3}$$

Here $(\Omega h^2)_{\text{min}}$ is set equal to 0.03.

For the neutralino relic abundance $\Omega_\chi h^2$ we employ the results of Ref.5. In Fig.2 we report $\Omega_\chi h^2$ as a function of $m_\chi$ for three representative neutralino compositions: i) a gaugino–dominated composition ($P = 0.9$), ii) a composition of maximal gaugino–higgsino mixing ($P = 0.5$), iii) a higgsino-dominated composition ($P = 0.1$). As expected, out of the three compositions displayed in Fig.2 the gaugino–dominated state provides the largest values of $\Omega_\chi h^2$. In order to have more substantial values of $\Omega_\chi h^2$, one has to consider more pure gaugino compositions ($P \gtrsim 0.99$). This is explicitly shown in Fig.3, where together with the reference value $P = 0.5$ also displayed are the rather extreme cases: $P = 0.01$ (very pure higgsino composition) and $P = 0.99$ (very pure gaugino composition). The very pronounced dips in the plots of Fig.s 2–3 are due to the s-poles in the $\chi - \chi$ annihilation cross section (exchange of the Z and of the Higgs neutral scalars).

In Fig.4 we display the values of $\Omega_\chi h^2$ versus $m_\chi$ in the form of a scatter plot obtained by varying $M_2$ and $\mu$ over a grid of constant spacing in the log-log plane of Fig.1.

By comparing Fig.4 with Fig.3 one sees that, at fixed $m_\chi$, the minimum of $\Omega_\chi h^2$ is provided by the $\chi$–configuration of maximal mixing, as is expected; furthermore, one sees that the large spread in values for $\Omega_\chi h^2$, displayed in Fig.4, is due to configurations of extremely pure composition.

## 3   Looking down, watching at muons that come up

Let us turn now to the indirect search for neutralino dark matter which can be performed by means of neutrino telescopes [8]. Neutralinos, if present in our galactic halo as dark matter components, would be slowed down by elastic scattering off the nuclei of the celestial bodies (Sun and Earth) and then gravitationally trapped inside them. Due to the process of neutralino capture these macroscopic bodies could accumulate neutralinos which would subsequently annihilate in pairs. An important outcome of this $\chi$–$\chi$ annihilation would be a steady flux of neutrinos from these celestial bodies.

The differential neutrino flux at a distance $d$ from the annihilation region is given by

$$\frac{dN_\nu}{dE_\nu} = \frac{\Gamma_A}{4\pi d^2} \sum_{F,f} B^{(F)}_{\chi f} \frac{dN_{f\nu}}{dE_\nu} \tag{4}$$

where $\Gamma_A$ is the annihilation rate and $F$ denotes the $\chi$–$\chi$ annihilation final states which are: 1) fermion–antifermion pairs, 2) pairs of neutral and charged Higgs bosons, 3) one gauge boson–one Higgs boson pairs, 4) pairs of gauge bosons; $B^{(F)}_{\chi f}$ denotes the branching ratio into the fermion $f$ (heavy quark or $\tau$ lepton), in the channel $F$; $dN_{f\nu}/dE_\nu$ denotes the differential distribution of the neutrinos generated by the semileptonic decays of the fermion $f$. The $\nu_\mu$'s, crossing the Earth, would convert into muons and generate a signal of up–going muons inside a neutrino telescope. Calculations of this muon flux from the original neutrino flux may be performed using standard procedures [8].

Particular care has to be taken in the evaluation of the annihilation rate $\Gamma_A$. This quantity is given by [9]

$$\Gamma_A = \frac{C}{2} \tanh^2\left(\frac{t}{\tau_A}\right) \tag{5}$$

where t is the age of the macroscopic body ($t = 4.5$ Gyr for Sun and Earth), $\tau_A = (CC_A)^{-1/2}$, $C$ is the capture rate of neutralinos in the macroscopic body and $C_A$ is the annihilation rate per effective volume of the body. The capture rate $C$ is provided by the formula [10]

$$C = \frac{\rho_\chi}{v_\chi} \sum_i \frac{\sigma_{\text{el},i}}{m_\chi m_i} (M_B f_i) \langle v_{esc}^2 \rangle_i X_i, \qquad (6)$$

where $\rho_\chi$ and $v_\chi$ are the neutralino local density and mean velocity, $\sigma_{\text{el},i}$ is the cross section of the neutralino elastic scattering off the nucleus $i$ of mass $m_i$ (for some properties of the elastic $\chi$–nucleus cross section see next Sect. 4 and Ref. [11]), $M_B f_i$ is the total mass of the element $i$ in the body of mass $M_B$, $\langle v_{esc}^2 \rangle_i$ is the square escape velocity averaged over the distribution of the element $i$, $X_i$ is a factor which takes account of kinematical properties occurring in the neutralino–nucleus interactions.

$C_A$ is given by [9]

$$C_A = \frac{<\sigma v>}{V_0} \left(\frac{m_\chi}{20 \text{ GeV}}\right)^{3/2} \qquad (7)$$

where $\sigma$ is the neutralino–neutralino annihilation cross section and $v$ is the relative velocity. $V_0$ is defined as $V_0 = (3 m_{Pl}^2 T / (2\rho \times 10 \text{ GeV}))^{3/2}$ where $T$ and $\rho$ are the central temperature and the central density of the celestial body. For the Earth ($T = 6000$ K, $\rho = 13$ g·cm$^{-3}$) $V_0 = 2.3 \times 10^{25}$ cm$^3$, for the Sun ($T = 1.4 \times 10^7$ K, $\rho = 150$ g·cm$^{-3}$) $V_0 = 6.6 \times 10^{28}$ cm$^3$.

For the computation of the capture rate (and then also of $\tau_A$) one has to use a specific value for the neutralino density $\rho_\chi$. Here, for any point of the model parameter space, we evaluate $\rho_\chi$ as explained in Sect.2.

In Fig.5 we give the results of our calculations for $\tanh^2(t/\tau_A)$ and for $\Gamma_A$, which include the rescaling for $\rho_\chi$. This figure refers to the Earth for the case of the usual three representative $\chi$–compositions. For simplicity, in this figure, as well as in the following ones, only the results concerning positive values of the parameter $\mu$ are shown; similar results hold for negative values of $\mu$. One clearly sees that equilibrium is not reached for $m_\chi \gtrsim m_W$, because of the substantial suppression introduced in $\Gamma_A$ by the factor $\tanh^2(t/\tau_A)$. In Fig.6 we show $\tanh^2(t/\tau_A)$ and for $\Gamma_A$ for the Sun; here the equilibrium between capture and annihilation is reached over the whole $m_\chi$ range.

Now we report some of our results about the flux of the up-going muons in the case of $\chi$–$\chi$ annihilation in the Earth. In Fig.s 7–8 we show the fluxes of the up–going muons as functions of $m_\chi$ for a number of values of the neutralino composition $P$, for $\chi$–$\chi$ annihilation in the Earth. The threshold for the muon energy is $E_\mu^{\text{th}} = 2$ GeV. We recall here that the present experimental upper bound for signals coming from the Earth is $4.0 \cdot 10^{-14}$cm$^{-2}$s$^{-1}$ (90 % C.L.) [12]. By comparing this upper limit with our fluxes we see that the regions explored by Kamiokande (at our representative point: $\tan\beta = 8, m_h = 50$ GeV) concern the mass range 50 GeV $\lesssim m_\chi \lesssim$ 65 GeV. These regions are illustrated in Fig.9 in a $\mu$–$M_2$ plot. In this figure we also display the regions which could be explored by a neutrino telescope with an improvement factor of 10 (and of 100) in sensitivity.

The location and the shape of the most easily explorable regions in the

$M_2 - \mu$ plane depend on the Earth chemical composition and on the neutralino composition in terms of the gaugino, higgsino components. In fact the capture rate of neutralinos is more effective when neutralino mass matches the mass of some of the main components of the Earth and when the neutralino is a large gaugino–higgsino mixture. Because of these two properties the signal is maximal along iso–mass lines in the range 50–65 GeV, with elongations along iso–composition lines of sizeable mixing.

In Fig.10 we report the fluxes for up–going muons as functions of $m_\chi$ due to $\chi$–$\chi$ annihilation in the Sun. As before the threshold for the muon energy is $E_\mu^{\rm th} = 2$ GeV. The evaluated fluxes are below the present experimental upper limit of Kamiokande: $6.6 \cdot 10^{-14} {\rm cm}^{-2} {\rm s}^{-1}$ (90 % C.L.) [12].

From these results it can be concluded that neutrino telescopes with an area above $10^5$ m$^2$ are very powerful tools for investigating neutralino dark matter in large regions of the parameter space. It also emerges from the previous results that the signals from the Earth and from the Sun somewhat complement each other to allow an exploration about DM neutralino over a wide range of $m_\chi$. Taking into account the appropriate on–source duty factor for the Sun, it turns out that the signal from the Sun starts overcoming the one from the Earth at about $m_\chi \sim 200$ GeV. This occurs since, even if the Sun is mainly composed of very light elements, its gravitational field is much more effective compared to the Earth's one in capturing neutralinos.

## 4  Patiently waiting that relic neutralinos hit our detector

Another way to search for dark matter neutralinos is the direct detection which relies on the measurement of the recoil energy of nuclei of a detector, due to elastic scattering of $\chi$'s. The relevant quantities to calculate are the differential rate (in the nuclear recoil energy $E_r$):

$$\frac{dR}{dE_r} = N_T \frac{\rho_\chi}{m_\chi} \int_{v_{min}(E_r)}^{v_{max}} dv f(v) v \frac{d\sigma_{el}}{dE_r}(v, E_r) \qquad (8)$$

and the integrated rate $R_{\rm int}$, which is the integral of Eq. (8) from the threshold energy $E_r^{\rm th}$, which is a characteristic feature of the detector, up to a maximal energy $E_r^{\rm max}$. In Eq.(8) $N_T$ denotes the number of target nuclei, $d\sigma_{\rm el}/dE_r$ is the differential elastic cross section and $f(v)$ is the distribution of $\chi$ velocities in the Galaxy. It is important to note again that the local density $\rho_\chi$ is evaluated here according to the procedure discussed in Sect. 2. In general, the $\chi$–nucleus cross section has two contributions: a coherent contribution, depending on $A^2$ ($A$ is the mass number of the nucleus) which is due to Higgs and $\tilde{q}$ exchange diagrams; a spin–dependent contribution, arising from $Z$ and $\tilde{q}$ exchange, proportional to $\lambda^2 J(J+1)$.

By way of example, let us remind the expression of the coherent cross section due to the Higgs–exchange [14]:

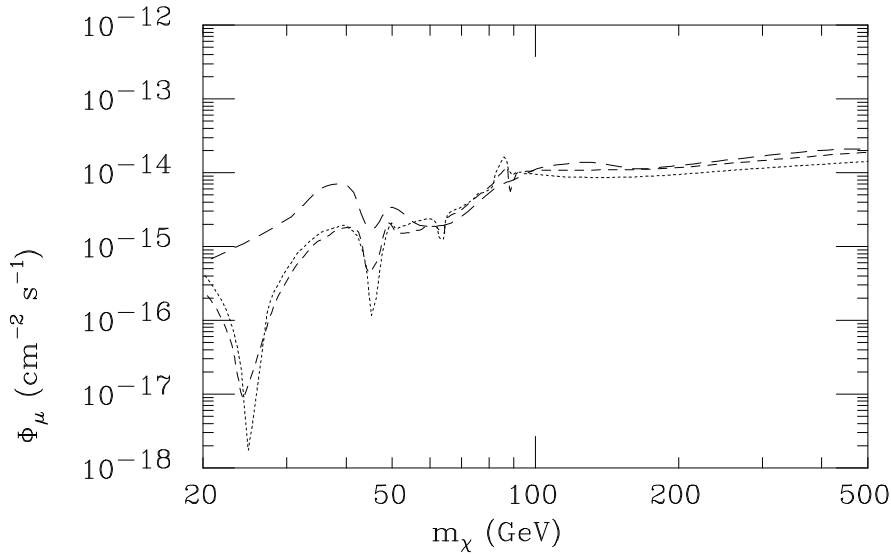

**Figure 10.** Fluxes of the up–going muons as functions of $m_\chi$ for $\chi$–$\chi$ annihilation in the Sun, for the three representative neutralino compositions $P = 0.1$ (dotted line), 0.5 (short–dashed line), 0.9 (long–dashed line). The threshold for the muon energy is $E_\mu^{\text{th}} = 2$ GeV.

$$\sigma_{\text{el},H} = \frac{8G_F^2}{\pi}\alpha_H^2 A^2 \frac{m_Z^2}{m_h^4}\frac{m_i^2 m_\chi^2}{(m_i^2 + m_\chi)^2} \qquad (9)$$

where $\alpha_H$ is a quantity depending on the neutralino–Higgs and the neutralino–quarks couplings. It is worth mentioning that $\alpha_H$ depends rather sensitively on the $\chi$–composition and on a number of parameters, such as $\tan\beta$ and the Higgs masses.

Except for very special points in the parameter space, the coherent contribution to elastic cross section strongly dominates over the spin–dependent one. For a detailed analysis on the calculation of the direct event rates see [15] and references quoted therein. For an experimental overview about dark matter detectors see Ref. [16].

Here, as an example, we simply report in Fig. 11 the event rates $R_{\text{int}}$ for a Germanium detector as a function of $m_\chi$ for neutralino compositions $P = 0.1, 0.5, 0.9$. Rates are calculated by integrating the differential rate of Eq.(8) over the electron–equivalent energy range (2–4) KeV. Fig. 12 shows $R_{\text{int}}$ for more pure neutralino compositions. In Fig. 13 we show the regions of the $M_2 - \mu$ parameter space which can be explored with an improvement of one and two orders of magnitude in the sensitivity of the detector.

As for the shape of these regions we refer to the comments presented above, in Sect. 3, in connection with Fig. 9. Again, the signals are higher along the iso–mass line with an $m_\chi$ matching the mass of any element composing the detector. Thus, using detectors of different compositions allows explorations of the $M_2 - \mu$ plane over a wide range in $m_\chi$. For instance, investigation of regions with small $m_\chi$ values (of order of 10 GeV) with very low threshold detectors [17] would be very interesting. In fact this $m_\chi$ range (which is excluded by accelerator data only under a number of assumptions) is out of reach for the indirect detection discussed in the previous Section.

## 5   An epilogue: how to save stationery and be happy

The procedure for getting reliable results about neutralino relic abundance and detection event rates requires rather elaborate calculations, demanding great care and accuracy. This is because in sizeable regions of the parameter space many different channels and final states are competing, and the coupling constants of the various processes are very sensitive to the values of the free parameters. Furthermore interplay between calculations of different quantities has to be taken into account, if one really cares about realistic evaluations; for instance, this is the case when a rescaled neutralino local density has to be used in the evaluation of the event rates.

It is clear that under simplifying assumptions and in special circumstances, for example when clear dominance of a particular process accidentally occurs, order–of–magnitude calculations may give reasonably good estimates. When this is the case, one may go through the straightforward exercise of feeding a few numbers in simple formulae (as the one in Eq.(9)) cheaply available in the market, and work out a reasonable (order of magnitude) result. It is obvious (but nevertheless worth mentioning) that these back–of–the–envelope estimates, very enjoyable for pedagogical purposes, may in no way be interpreted as substitute for realistic calculations. The inadequacy of back–of–the–envelope evaluations in this field may be proved quite easily, even on the back of a stamp if saving stationery and not caring about accuracy in calculations is the main point of the game!